\documentstyle[12pt,worldsci,epsfig]{article}
\newcommand{\meff}{M_{\mbox{\small eff}}}
\newcommand{\vbead}{V_{\mbox{\small J}}}
\newcommand{\thetarl}{\theta_{\rho\lambda}}
\newcommand{\br}{{\bf r}}
\newcommand{\brho} {{\mbox{\boldmath $\rho$}}}
\newcommand{\blambda} {{\mbox{\boldmath $\lambda$}}}
\newcommand{\beqn}{\begin{equation}}
\newcommand{\eeqn}{\end{equation}}

\begin{document}
\title{(HYBRID) BARYONS:  QUANTUM NUMBERS AND ADIABATIC POTENTIALS}
\author{Philip R. Page\thanks{prp@t5.lanl.gov. Work done in collaboration
with Simon Capstick. 
}\\
{\em Theoretical Division, MS B283, Los Alamos National Laboratory, \\
Los Alamos, NM 87545, USA}\\
\vspace{0.3cm}
}
\maketitle
\setlength{\baselineskip}{2.6ex}

\vspace{0.7cm}
\begin{abstract}

We construct (hybrid) baryons in the flux--tube model
of Isgur and Paton. In the limit of adiabatic quark motion, 
we build proper eigenstates of orbital angular momentum and
indicate the flavour, spin, chirality and $J^{P}$ of (hybrid) baryons. The adiabatic
potential is calculated as a function of the quark positions. 
\end{abstract}
\vspace{0.7cm}

\section{Introduction}

Hybrid baryons are bound states of three quarks with an explicit
excitation in the gluon field of QCD. 
The construction of (hybrid) baryons in a
model motivated from and consistent with 
lattice gauge theory, the non--relativistic
flux--tube model of Isgur and Paton, was detailed in ref. \cite{hadron} 
There we studied the detailed flux dynamics and built
the flux hamiltonian. 
A minimal amount of 
quark motion is allowed in response to flux motion, 
in order to work in the centre of mass frame.
Otherwise, we make the so--called ``adiabatic'' approximation, where the
flux motion adjusts itself instantaneously to the motion of the quarks.
The main result was that the lowest
flux excitation can to a high degree of accuracy (about 5\%) be simulated
by neglecting all flux--tube motions except the vibration of a 
junction. 
The junction acquires an effective mass from the motion of the 
remainder of the flux--tube and the quarks.
The model is then simple: a junction is 
connected via a linear potential to the three quarks. 
The ground state of the junction motion corresponds to a conventional
baryon and the various excited states to hybrid baryons.

\section{Quantum numbers of (hybrid) baryons}

\begin{table}[t]
\begin{center}
\caption{\small Quantum numbers of ground state  (hybrid) baryons for the  adiabatic surfaces $B,H_1,H_2$ and $H_3$. Here $B$ denotes the conventional baryon. The mass ordering is $B < H_1 < H_2 < H_3$. In the absense of spin dependent forces all ground states corresponding to a given adiabatic surface are degenerate. The quantum number notation is $(N,\Delta)^{2S+1}J^P$, where $N,\Delta$ is the flavour structure of the wave function (i.e. those of the conventional baryons $N,\Delta$ respectively) and $P$ the parity. } 
\label{tabqu}
\begin{tabular}{|c||r|l|l|}
\hline 
(Hybrid) Baryon          & Chirality & $L$ & $(N,\Delta)^{2S+1}J^P$ \\
\hline 
$B       $ & 1 & 0 &  $N^2 {\frac{1}{2}}^+, \; \Delta^4 {\frac{3}{2}}^+$\\
$H_1^S   $ & 1 & 1 &  $N^2 {\frac{1}{2}}^+, \; N^2 {\frac{3}{2}}^+, \; \Delta^4 {\frac{1}{2}}^+, \; \Delta^4 {\frac{3}{2}}^+, \; \Delta^4 {\frac{5}{2}}^+$\\
$H_1^A   $ & 1 & 1 &  $N^2 {\frac{1}{2}}^+, \; N^2 {\frac{3}{2}}^+$\\
$H_2^S   $ & 1 & 1 &  $N^2 {\frac{1}{2}}^+, \; N^2 {\frac{3}{2}}^+, \; \Delta^4 {\frac{1}{2}}^+, \; \Delta^4 {\frac{3}{2}}^+, \; \Delta^4 {\frac{5}{2}}^+$\\
$H_2^A   $ & 1 & 1 &  $N^2 {\frac{1}{2}}^+, \; N^2 {\frac{3}{2}}^+$\\
$H_3^S   $ &-1 & 0 &  $N^2 {\frac{1}{2}}^-, \; \Delta^4 {\frac{3}{2}}^-$\\
$H_3^A   $ &-1 & 0 &  $N^2 {\frac{1}{2}}^-$\\
\hline 
\end{tabular}
\end{center}
\end{table}

The junction can move in three directions, and correspondingly be excited
in three ways, giving the hybrid baryons $H_1, H_2$ and $ H_3$. For each junction
excitation, it is found that the junction wave function can be realized
to be either totally symmetric, or totally antisymmetric under quark label
exchange, indicated by $H^S$ and $H^A$ respectively.

The quantum numbers of the lowest--lying states that can be constructed
on the adiabatic surfaces corresponding to each of the six hybrid baryons
are indicated in Table \ref{tabqu}.

Since quarks are fermions, the wave function 
should be totally antisymmetric under quark label exchange, called
the Pauli principle. The colour structure of hybrid baryons are taken to
be identical to those of conventional baryons, i.e. it is totally antisymmetric
under label exchange. This imposes constraints on the combination of 
flavour and non--relativistic spin $S$ of the three quarks that is allowed.
The combinations are indicated in Table \ref{tabqu}.

``Chirality'' gives the behaviour of the junction wave function  under reflection in the plane spanned by the quarks, when the positions of the quarks are fixed.

Let $L$ be the orbital angular momentum of the quarks and the junction. 
It is possible to argue that  $L=1$ for the ground state
$H_1$ and $H_2$ hybrid baryons, while $L=0$ for the ground state conventional
and $H_3$ hybrid baryons.
The total angular momentum ${\bf J} = {\bf L} + {\bf S}$. Since $L=0$ for
ground state conventional and $H_3$ hybrid baryons, $J=S$. Since $L=1$ for   
ground state $H_1, H_2$ hybrid baryons, $J=\frac{1}{2},\frac{3}{2}$ for $S=\frac{1}{2}$,
and $J=\frac{1}{2},\frac{3}{2},\frac{5}{2}$ for $S=\frac{3}{2}$.
These assignments are indicated in Table \ref{tabqu}.

\section{The potential in which the quarks move}

\begin{figure}[t]
\vspace{-1.6cm}
\begin{centering}
\epsfig{file=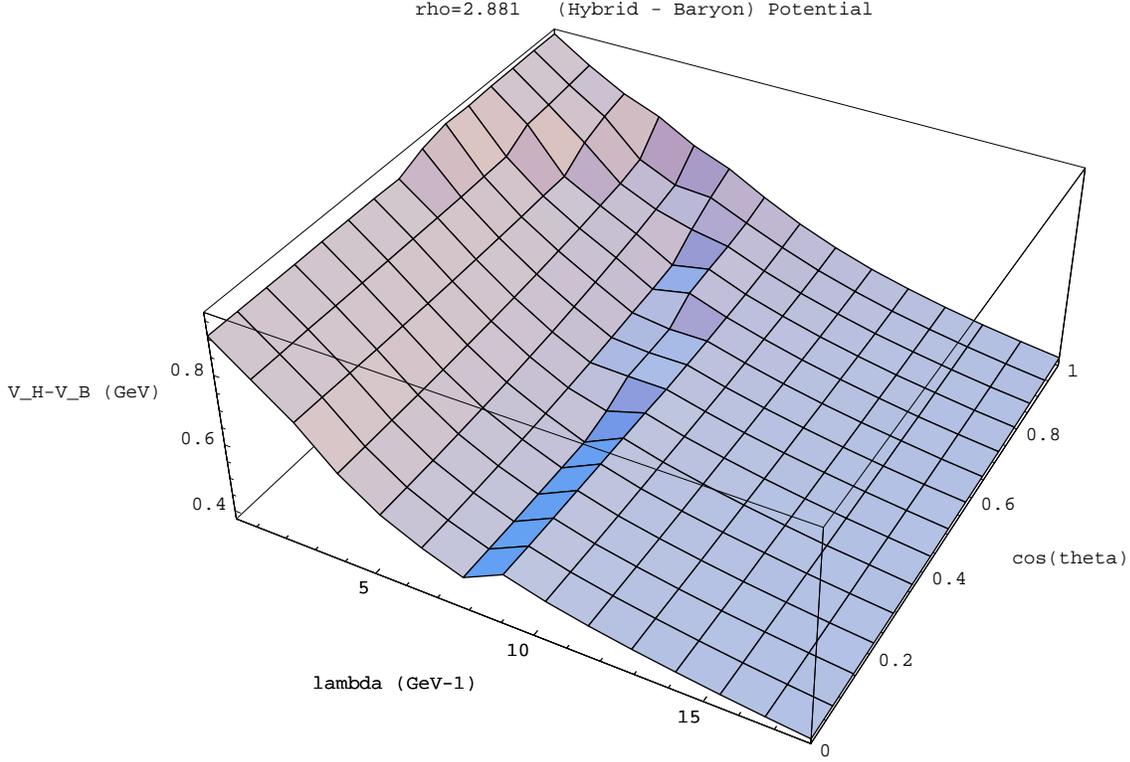,width=15cm,angle=0}
\vspace{-2.5cm}
\caption[x]{Difference of the hybrid and conventional baryon potentials
(in GeV) for $\rho=2.881$ GeV$^{-1}$ as a function of $\lambda$ (in GeV$^{-1}$) and
$\cos\thetarl$.}
  \label{fig:TheLoop}
  \end{centering}
\end{figure}
\vspace{0.3cm}

We shall now calculate the junction energy (``adiabatic potential'')
 for the ground and first 
excited states of the junction as a function of the quark positions.

Define

\beqn\label{rho}
{\brho} = \frac{\br_1 - \br_2}{\sqrt{2}}\hspace{1cm}
{\blambda} = \frac{\br_1+\br_2-2 \br_3}{\sqrt{6}}\hspace{1cm}
\cos \thetarl = \frac{\brho\cdot\blambda}{\rho\lambda}
\eeqn
where $\br_i$ denotes the positions of the quarks. The energy is a 
function of $\rho$, $\lambda$ and $\thetarl$.

The procedure for evaluating the conventional baryon potential
is as follows. 
We numerically evaluate $V_B (l_1,l_2,l_3)$ 
by solving the Schr\"{o}dinger Equation for the junction hamiltonian,
$(\frac{1}{2}\meff^{\infty}{\bf
\dot{r}^2}+\vbead)\Psi_B(\br) =V_B (l_1,l_2,l_3)\Psi_B(\br)  $,
variationally using an ansatz ground state simple harmonic oscillator 
wave function. $\meff^{\infty}$ is the effective mass of the junction in the
limit where the flux--tubes between the junction and quarks are continuous
strings. $\vbead$ is the linear potential between the junction and the
three quarks. The hybrid baryon $H_1$
potential $V_{H_1} (l_1,l_2,l_3)$ is solved using 
$(\frac{1}{2}\meff^{\infty}{\bf
\dot{r}^2}+\vbead)\Psi_{H_1}(\br) =V_{H_1} (l_1,l_2,l_3)\Psi_{H_1}(\br)$ 
with a first excited state simple harmonic oscillator 
wave function as an ansatz. 

\begin{figure}[t]
\vspace{-1.8cm}
\begin{centering}
\epsfig{file=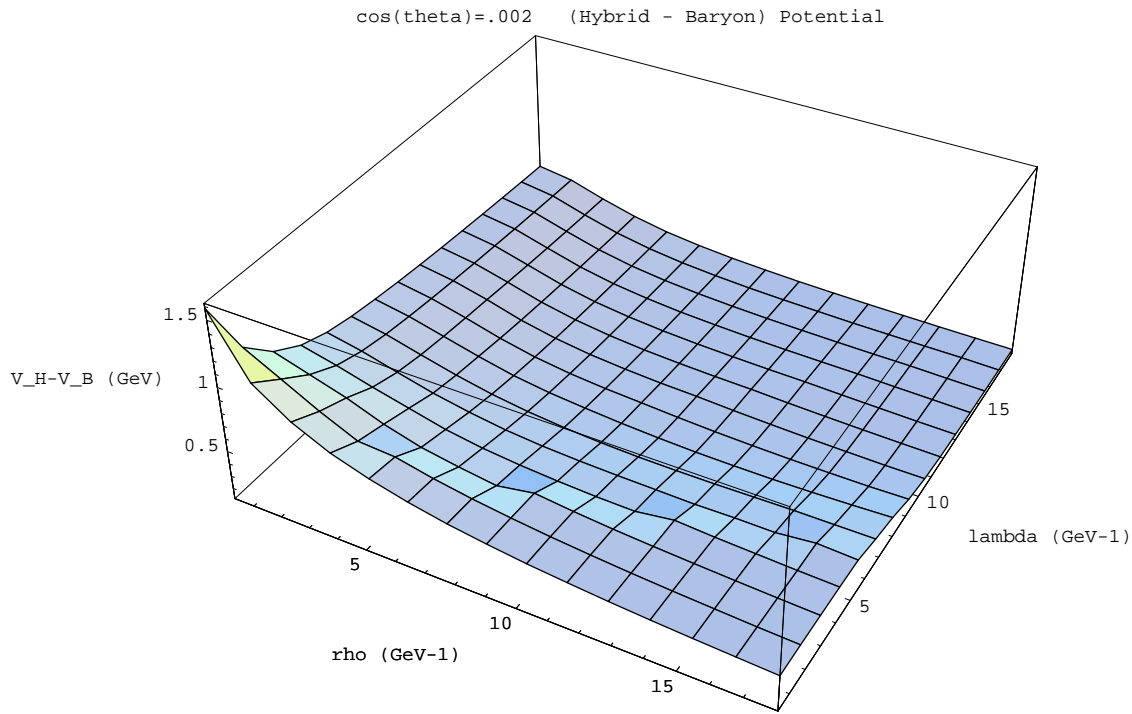,width=15cm,angle=0}
\vspace{-2.8cm}
\caption[x]{Difference of the hybrid and conventional baryon potentials
for $\cos\thetarl=0.002$ as a function of $\rho$ and $\lambda$ 
(in GeV$^{-1}$).}
  \label{fig:TheLoop1}
  \end{centering}
\end{figure}

The difference between the hybrid and conventional baryon potentials
 is plotted in Figures \ref{fig:TheLoop} - \ref{fig:TheLoop1}. Since
the potentials are functions of $\rho,\lambda$ and $\thetarl$, one of the
variables is held fixed at a typical value for clarity of presentation.

The (hybrid) baryon potential can be seen to increase
 when $\rho\lambda$ is small.
Numerically, the ratio of the hybrid to baryon potential 
is found to be $1.44 - 1.6$ for all $\rho,\lambda$ and $\thetarl$.

A preliminary estimate of the ground state $H_1$ hybrid baryon mass of
$\sim 2$ GeV has also been made by adding the 
difference between the hybrid and conventional baryon potentials to the
phenomenologically successful baryon potential used in ref. \cite{capstick86}

\section{Conclusions}

The spin and flavour structure of the six hybrid baryons have been specified.
Exchange symmetry constrains the spin and flavour of the (hybrid) baryon 
wave function. The orbital angular momentum of the low--lying hybrid baryon
is argued to be unity. The adiabatic potentials have been calculated numerically.
The low--lying hybrid baryon mass has been estimated numerically. 


\end{document}